\documentclass[onecolumn,prc,superscriptaddress,unsortedaddress,preprintnumbers,amsmath,amssymb]{revtex4}
\usepackage{graphicx}% Include figure filesph
\usepackage{dcolumn}% Align table columns on decimal point
\usepackage{txfonts}
\usepackage{bm}% bold math
\usepackage[dvipdfmx,bookmarks=true,colorlinks,%
citecolor=blue,linkcolor=blue,anchorcolor=blue,filecolor=blue,urlcolor=blue,%
           ]{hyperref}
\def\beq{\begin{equation}}
\def\eeq{\end{equation}}
\def\bea{\begin{eqnarray}}
\def\eea{\end{eqnarray}}

\def\fun#1#2{\lower3.6pt\vbox{\baselineskip0pt\lineskip.9pt
  \ialign{$\mathsurround=0pt#1\hfil##\hfil$\crcr#2\crcr\sim\crcr}}}
  
\begin{document}
%\begin{CJK*} {GBK} {song}
\preprint{}

\title{$r$-mode instability of neutron stars in Low-mass X-ray binaries: effects of Fermi surface depletion and superfluidity of dense matter}

\author{J. M. Dong}\affiliation{Institute of Modern Physics, Chinese
Academy of Sciences, Lanzhou 730000, China} \affiliation{School of
Physics, University of Chinese Academy of Sciences, Beijing 100049,
China}
\date{\today}

\begin{abstract}
The nucleon-nucleon correlation between nucleons leads to the Fermi
surface depletion measured by a $Z$-factor in momentum distribution
of dense nuclear matter. The roles of the Fermi surface depletion
effect ($Z$-factor effect) and its quenched neutron triplet superfluidity of nuclear matter in
viscosity and hence in the gravitational-wave-driven $r$-mode instability of neutron stars (NSs)
are investigated. The bulk viscosity is reduced by both the two
effects, especially the superfluid effect at low temperatures which
is also able to reduce the inferred core temperature of NSs.
Intriguingly, due to the neutron superfluidity, the core temperature
of the NSs in known low-mass X-ray binaries (LMXBs) are found to be
clearly divided into two groups: high and low temperatures which
correspond to NSs with short and long recurrence times for
nuclear-powered bursts respectively. Yet, a large number of NSs in
these LMXBs are still located in the $r$-mode instability region. If
the density-dependent symmetry energy is stiff enough, the occurence
of direct Urca process reduces the inferred core temperature by
about one order of magnitude. Accordingly, the contradiction between
the predictions and observations is alleviated to some extent, but
some NSs are still located inside the unstable region.

\end{abstract}

\maketitle
\maketitle

\section{Introduction}\label{intro}\noindent

A great deal of attention has been paid to gravitational waves
following its discovery from binary black holes merge (Abbott et al.
2016) and binary neutron stars (NSs) merge (Abbott et al. 2017). As
a species of compact objects, NSs themselves can radiate
gravitational waves due to for example the magnetic deformation
(Bonazzola \& Gourgoulhon 1996; Regimbau \& de Freitas Pacheco 2001;
Stella et al. 2005; Dall 'Osso, Shore \& Stella 2009; Marassi et al.
2011, Cheng et al. 2015, 2017) and $r$-mode instability (Andersson
1998). The $r$ mode is a class of fluid mode of oscillation with
Coriolis force as restoring force which is analogous to Earth's
Rossby waves, leading to the emission of gravitational waves in hot
and rapidly rotating NSs due to the Chandrasekhar-Friedmann-Schutz
instability, and hence it prevents the pulsars (rotating NSs) from
reaching their Kepler rotational frequency $\Omega_{\text{Kepler}}$.
The emission of gravitational waves is able to excite $r$ modes in NS
core in turn and causes the oscillation amplitude to grow, hence
resulting in a positive feedback. Such gravitational radiation
induced by the $r$-mode instability is perhaps detectable with
ground-based instruments in the coming years, and thus potentially
provides a probe to uncover the interior information of NSs. On the
other hand, the temperature-dependent damping mechanisms attributed
to the bulk and shear viscosities, hinder the growth of the $r$
mode. The $r$-mode instability window borders a critical curve that
is determined by the balance of evolution time scales ($r$-mode
driving and viscosity damping time scales are equal) in the
frequency-temperature ($\nu_s-T$) plane. Above this critical curve
the $r$-mode instability is active. In other words, the unstable
window depends on the competition between the gravitational
radiation and the viscous dissipation (see e.g. Andersson \&
Kokkotas 2001, Haskell 2015, Kostas et al. 2016, for review).

The low mass X-ray binary (LMXB) is a binary system where a compact
object such as a NS (We discuss this case in the present
work) is accreting matter from its low-mass companion that fills its
Roche lobe. Ho et al. (2011) assumed that the spin-up torque from
accretion is balanced by the spin-down torque from gravitational
radiation due to the unstable $r$-mode, and the associated heating
is equal to cooling via neutrino emissions, and therefore concluded
that many NSs are located in the $r$-mode instability region. Since
the $r$-mode instability limits the spin-up of accretion powered
millisecond pulsars in LMXBs, the rapidly rotating pulsars, such as
the PSR J1748-2446ad rotating at 716 Hz, are difficult to understand.
A better understanding of relevant damping mechanisms is
particularly necessary.

Bulk viscosity appears, if the perturbations in pressure and density
induced by the $r$-mode oscillations drives the dense matter away
from $\beta$-equilibrium. Consequently, energy is dissipated as the
system tries to restore its chemical equilibrium because of weak
interaction. The bulk viscosity due to the modified Urca reactions
(or perhaps the direct Urca for large mass NSs) provides the
dominant dissipation mechanism at high temperatures ($\gtrsim
10^{9}$K). However, at low temperatures ($\lesssim 10^{9}$K), the
shear viscosity caused by the neutron scattering and electron
scattering or crust-core interface, is the primary mechanism for the
damping of $r$ modes. The viscosity is expected to be affected
significantly by the superfluidity.

Superfluidity is an intriguing feature of dense nuclear matter,
which receives great interest since it plays an essential role in
the NS thermal evolution (Yakovlev et al. 1999, 2001; Page et al.
2004, 2006). Dong et al. (2013, 2016) found that the neutron
$^3PF_2$ superfluidity for pure neutron matter or $\beta$-stable
nuclear matter is strongly reduced by about one order of magnitude by the
Fermi surface depletion effect (i.e., $Z$-factor effect) with the
help of the microscopic Brueckner theory starting from bare
nucleon-nucleon interactions. The nucleon-nucleon correlation, in particular the short-range correlation including
short-range repulsion and tensor interaction, is so strong that it creats a high-momentum tail,
giving rise to the $Z$-factor. The $Z$-factor at the Fermi surface
is equal to the discontinuity of the occupation number in momentum
distribution according to the Migdal-Luttinger theorem (Migdal
1957), as shown in the inset of Fig.~\ref{fig:gap}(a). Therefore, it
characterizes the deviation from the right-angle momentum
distribution of perfect degenerate Fermi gas at zero-temperature,
and hinders the particle transitions around the Fermi surface which
could affect many properties of fermion systems related to
particle-hole excitations. For instance, the $Z$-factor has
far-reaching impact on the nuclear structure (Subedi et al. 2008;
Hen et al. 2014), superfluidity of dense nuclear matter (Dong et al.
2013, 2016), NS thermal evolution (Dong et al. 2016), and the
European Muon Collaboration effect (Hen et al. 2017; Duer et al.
2018), highlighting its fundamental importance. In the present work,
the influences of the $Z$-factor along with its quenched
superfluidity on the $r$-mode instability of NSs in LMXBs are
investigated in detail.

\section{Effects of the $Z$-factor and its quenched superfluidity on viscosity}\label{intro}\noindent
In addition to depressing the superfluidity, the $Z$-factor effect
is also able to change the viscosity directly because of the
particle number depletion at the Fermi surface. For the neutron and
proton $Z$-factors at the Fermi surfaces of $\beta$-stable NS matter
at different nucleon number density $\rho$, we use simple formulas that
depend on several parameters to fit the results of Dong et al.
(2016) which is achieved from the Brueckner theory using the Argonne
V18 nucleon-nucleon interaction plus a microscopic three-body force,
given by
\begin{eqnarray}
Z_{F,n}(\rho ) &=&0.907-0.233\rho -0.480\rho ^{2}+0.481\rho ^{3}, \nonumber\\
Z_{F,p}(\rho ) &=&%
\begin{cases}
0.351+2.332\rho ,\quad \rho \leq 0.15\text{ fm}^{-3} \\
0.656+0.451\rho -1.151\rho ^{2}+0.576\rho ^{3},\quad \rho >0.15\text{ fm}%
^{-3}%
\end{cases} \label{AA}
\end{eqnarray}
for the sake of application, where the fraction of each component
for $\beta$-stable matter is determined by the well-known
variational APR equation of state (EOS) (Akmal et al. 1998). The
non-rotating NS maximum mass of $2.2 M_{\odot}$, the canonical NS
radius of $11.6$ km, and stellar thermal evolution obtained from
this APR EOS are compatible with astrophysical observations (Page et
al. 2004), and it is one of the most popular EOS to be employed to
study the NS interior physics. The isospin-dependent part of EOS,
i.e., the symmetry energy, from the Brueckner-Hartree-Fock approach
is so stiff that the direct Urca (DUrca) reaction occurs even in
$1.2 M_{\odot}$ low-mass NSs (Yin \& Zuo 2013), which is not
consistent with the current understanding that the DUrca process
does not occur in $1.4 M_{\odot}$ canonical NSs (Lattimer \& Prakash
2004; Page et al. 2004; Brown et al. 2018). Fortunately, the
$Z$-factor-quenched superfluidity is so weak that the energy
gap is not so sensitive to the nucleon-nucleon interaction and
single-particle potential any longer, which is beneficial to obtain
reliable superfluid gaps. For example, the inclusion of three-body
force does not change the weak neutron $^3PF_2$ superfluidity gap
very much (Dong et al. 2013). The $Z_F$ itself of neutron-rich
matter is not very sensitive to the isospin asymmetry
$\beta$ at densities of interest if $\beta >0.5$ (Yin et al. 2013).
Therefore, the inconsistent treatment here is not expected to change
the final conclusions. Because the Fermi surface depletion hinders
particle-hole excitation around the Fermi level, the bulk viscosity
$\xi$, being related to neutrino emission, is expected to be reduced
with the inclusion of the $Z$-factor. The momentum distribution
function $n(k)$ of nucleons near the Fermi surface at finite
temperature $T$ (not quite high) is given by (Dong et al. 2016)
\begin{equation}
n(k)=\frac{Z_{F}}{1+\exp \left( \frac{\omega -\mu }{T}\right)
},\quad k\approx k_{F}, \label{Occupation}
\end{equation}
where $\omega$ is single-particle energy, and $\mu$ the chemical
potential. The nucleon-nucleon correlation between nucleons quenches the
occupation probability by a factor of $Z_F$ at the Fermi surface.

%%%%%%%%%%%%%%%%%%%%%%%%%%%%%%%%%%%%%%%%%%%%%%%%%%%%%%%%%%%%%%%%
\begin{figure}[htbp]
\begin{center}
\includegraphics[width=0.55\textwidth]{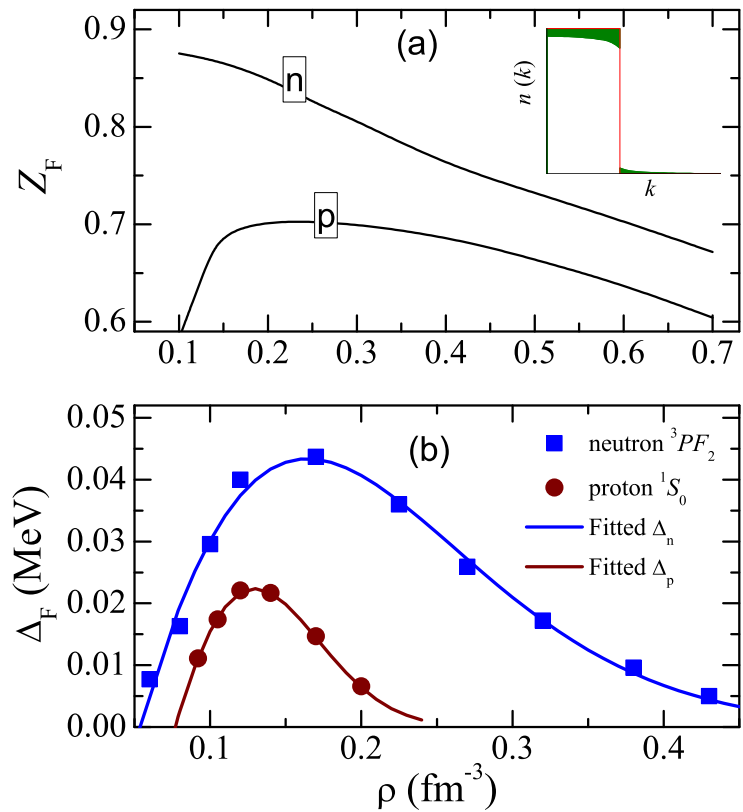}
\caption{(Color online) (a): $Z$-factor at the Fermi surface as a
function of nucleon number density $\rho$ in $\beta$-stable NS matter (Dong
et al. 2016). The inset shows a schematic illustration of the
momentum distribution due to the $Z$-factor effect. (b): neutron
$^3PF_2$ superfluid gap VS. nucleonic density in $\beta$-stable
matter (Dong et al. 2016; Li et al. 2020), compared with the fitting
results that denoted by the solid curves.}\label{fig:gap}
\end{center}
\end{figure}
%%%%%%%%%%%%%%%%%%%%%%%%%%%%%%%%%%%%%%%%%%%%%%%%%%%%%%%%%%%%%%%%%%%%%%%

The rapid cooling of the NS in Cas A has been revealed through the
ten-year observations (Heinke \& Ho 2010), which helps one to
extract the interior information of NSs (Page et al. 2011; Shternin
et al. 2011; Blaschke et al. 2012, 2013; Sedrakian 2013; Newton et
al. 2013; Bonanno et al. 2014; Ho et al. 2015). Then combined with
theoretical analysis, Page et al. (2011) claimed that this is the
first direct evidence that neutron triplet superfluidity and proton
singlet superconductivity occur at supranuclear densities within
NSs, with the critical temperature for neutron $^3P_2$ superfluidity
$T_{\text{cn,max}}=5\times10^8$ K (the corresponding superfluid
gap 0.08 MeV). However, Posselt et al. reported that a statistically
significant temperature drop for the NS in Cas A is not present.
Therefore, a reliable calculation of the superfluid gap
theoretically is especially necessary. When the $Z$-factor effect is
taken into account, the peak-value of neutron $^3PF_2$ superfluid gap
is about 0.04 MeV for pure neutron matter as well as $\beta$-stable
NS matter (Dong et al., 2013, 2016), in agreement with
other predictions (Ding et al. 2016) but much lower than the
constrained value of 0.08 MeV by Page et al. (2011). The
superfluid gap quenches all processes that involve elementary
excitations around the Fermi surface, leading to remarkable effects
on the neutrino emissivity, heat capacity, thermal conductivity, and
hence NS thermal evolution. Since the core temperatures
for NSs in LMXBs were inferred to be $(1\sim3)\times 10^{8}$K (Ho et
al. 2011) which is basically below the critical temperature of
neutron $^3PF_2$ superfluidity, such a superfluidity is expected to
affect the $r$-mode instability distinctly.

For the angular-averaged neutron $^3PF_2$ superfluid gap and proton
$^1S_0$ superconducting gap quenched by the $Z$-factor effect at
zero-temperature in $\beta$-stable matter, we fit the results of
Dong et al. (2016) and Dong (2019) obtained with the generalized
Bardeen-Cooper-Schriffer method combined with the Brueckner theory,
as summarized in Fig.~\ref{fig:gap} (b), which take the form of
\begin{eqnarray}
\Delta _{n}(\rho ) &=&(0.943\rho -0.050)\exp \left[ -\left( \frac{\rho }{%
0.177}\right) ^{1.665}\right] ,   \label{gap} \\
\Delta _{p}(\rho ) &=&(1.015\rho -0.078)\exp \left[ -\left( \frac{\rho }{%
0.136}\right) ^{2.823}\right] ,
\end{eqnarray}
as a function of nucleon number density $\rho$ and the corresponding critical
temperature is $T_c=0.57\Delta$ (Page et al. 2004). The results
perhaps are also useful to help one to understand pulsar glitches.
The proton $^1S_0$ gap is much smaller than the neutron $^3PF_2$
gap, and its superfluid domain is rather narrow. In addition, the
proton fraction is much smaller than the neutron one for
$\beta$-stable NS matter. Therefore, the proton superconductivity
will not be considered in the following discussion.

We explore the influence of the $Z$-factor and superfluidity on the
viscosity. The bulk viscosity of $npe\mu$ matter is mainly
determined by the reaction of DUrca process $n\rightarrow
p+l+\overline{\nu _{l}}, p+l\rightarrow n+\nu _{l}$ together with
the modified Urca (MUrca) processes $n+N\rightarrow
p+N+l+\overline{\nu _{l}},p+N+l\rightarrow n+N+\nu _{l}$, where $N$
denotes neutron ($n$) or proton ($p$) and $l$ denotes electron ($e$)
or muon ($\mu$). The DUrca process is most efficient for neutrino
($\nu$) emission, but it occurs only if the proton fraction is high
enough to reach a threshold. Frankfurt et al. (2008) concluded that
the high-momentum tail of the nucleon momentum distribution induced
by the short-range correlation leads to a significant enhancement of
the neutrino emissivity of the DUrca process, and this rapid process
can be opened even at low proton fraction. However, Dong et al.
(2016) gave the opposite conclusions, that is, the neutrino
emissivity is reduced instead of enhanced, and the threshold
condition for the DUrca reaction is almost unchanged. The
$\beta$-decay of neutrons and its inverse process can be cyclically
driven just by thermal excitations, and high momentum tail cannot
participate in the DUrca process and also the MUrca processes. The
explicit discussion are presented in Dong et al. (2016).

The partial bulk viscosity of $npe\mu$ matter induced by a
non-equilibrium DUrca process without the $Z$-factor effect ($Z=1$)
is written as (Haensel et al. 2000)
\begin{eqnarray}
\xi _{l,0}^{(D)} &=&K\int_{0}^{\infty }dx_{\nu }x_{\nu }^{2}\int
dx_{n}dx_{p}dx_{l}\bigg\{f(x_{n})f(x_{p})f(x_{l})  \notag \\
&&\cdot \left[ \delta (x_{n}+x_{p}+x_{l}-x_{\nu }+\zeta )-\delta
(x_{n}+x_{p}+x_{l}-x_{\nu }-\zeta )\right] \bigg\},  \label{BB1}
\end{eqnarray}
through a complicated deduction, where $f(x)=1/(1+e^{x})$ in the
phase space integral is a Fermi-Dirac function and $K$ is related to
the temperature $T$, $r$-mode angular frequency $\omega$, nuclear
matter EOS and possible superfluid gap. $\zeta$ is a parameter to
measure the deviation of the system from $\beta$-equilibrium. Due to
the strong degeneracy of nucleons and electrons, the main
contribution to the above integral comes from the very narrow
regions of momentum space near the corresponding Fermi surfaces
$k_F$, just as the calculation of neutrino emissivity of the DUrca
process in Yakovlev et al. (2001). If the $Z$-factor effect is
included, the above Fermi-Dirac distribution $f(x)$ near the Fermi
surface for nucleons should be replaced by Eq. (\ref{Occupation}),
i.e., $Z_{F}/(1+e^{x})=Z_{F} f(x)$, resulting in an additional
factor $Z_{F,n}Z_{F,p}$ appearing in the right hand side of Eq.
(\ref{BB1}). Therefore, the bulk viscosity induced by DUrca reaction
is given by
\begin{equation}
\xi _{l}^{(D)}=Z_{F,n}Z_{F,p}\xi _{l,0}^{(D)}. \label{BB}
\end{equation}
The explicit expression of the $\xi _{l,0}^{(D)}$ takes the form of (Haensel et al. 2000)
\begin{eqnarray}
\xi _{l,0}^{(D)} &=&8.553\times 10^{24}\frac{m_{n}^{\ast }}{m_{n}}\frac{%
m_{p}^{\ast }}{m_{p}}\left( \frac{n_{e}}{0.16}\right) ^{1/3} \nonumber\\
&&\cdot \frac{T_{9}^{4}}{\omega _{4}^{2}}\left( \frac{C_{l}}{100\text{ MeV}}%
\right) ^{2}\Theta _{npl}\text{\ g cm}^{-1}\text{s}^{-1},
\end{eqnarray}
with $\omega _{4}=\omega /(10^{4}$s$^{-1})$, $T_{9}=T/(10^{9}$K$)$
and $C_{l}=4(1-2Y_{p})\rho dS(\rho )/d\rho
-k_{F,l}^{2}/[12(1-2Y_{p})S(\rho )]$ where $Y_{p}$ and $S(\rho )$
are the proton fraction and density-dependent symmetry energy
respectively. $n_e$ and $m^{\ast }$ are the electron number density
in units of fm$^{-3}$ and nucleonic effective mass, respectively.
The step function is $\Theta _{npl}=1$ if the DUrca process opens
for $k_{F,n}<(k_{F,l}+k_{F,p})$. The $\xi _{l,0}^{(D)}$ could
include the superfluid effect by a multiplying control function whose
tedious expression can be consulted in Haensel et al. (2000). The
angular frequency $\omega$ of the $r$-mode ($l=m=2$) in a corotating
frame is given by $\omega=2m\Omega/l(l+1)$ (Andersson 2001), where
$\Omega$ is the angular velocity of the rotating star. Similarly,
the distribution function $f(x_{n})f(x_{p})f(x_{N})f(x_{N^{\prime
}})$ that appears in phase space integral (Haensel et al. 2001) for
the bulk viscosity produced by the MUrca processes, is replaced by
$[Z_{F,n}f(x_{n})][Z_{F,p}f(x_{p})][Z_{F,N}f(x_{N})][Z_{F,N^{\prime
}}f(x_{N^{\prime }})]$, and thus we obtain
\begin{eqnarray}
\xi _{l}^{(Mn)} &=&Z_{F,n}^{3}Z_{F,p}\xi _{l,0}^{(Mn)}, \label{MU1}\\
\xi _{l}^{(Mp)} &=&Z_{F,n}Z_{F,p}^{3}\xi _{l,0}^{(Mp)}, \label{MU2}
\end{eqnarray}
where the superscript $N=n(p)$ denotes the neutron (proton) branch
of the MUrca processes. The $\xi _{l,0}^{(Mn)}$ and $\xi
_{l,0}^{(Mp)}$ are given by (Haensel et al. 2001)
\begin{eqnarray}
\xi _{e,0}^{(Mn)} &=&1.49\times 10^{19}\left( \frac{m_{n}^{\ast }}{m_{n}}%
\right) ^{3}\frac{m_{p}^{\ast }}{m_{p}}\left( \frac{n_{p}}{0.16}\right)
^{1/3} \nonumber \\
&&\cdot \frac{T_{9}^{6}}{\omega _{4}^{2}}\left( \frac{C_{e}}{100\text{ MeV}}%
\right) ^{2}\alpha _{n}\beta _{n}\text{\  g cm}^{-1}\text{s}^{-1}, \\
\xi _{e,0}^{(Mp)} &=&\xi _{e,0}^{(Mn)}\left( \frac{m_{p}^{\ast }}{%
m_{n}^{\ast }}\right) ^{2}\frac{(3k_{F,p}+k_{F,e}-k_{F,n})^{2}}{%
8k_{F,p}k_{F,e}}\Theta _{pe}, \\
\xi _{\mu,0}^{(Mn)} &=&\xi _{e,0}^{(Mn)}\left( \frac{k_{F,\mu }}{k_{F,e}}%
\right) \left( \frac{C_{\mu }}{C_{e}}\right) ^{2}, \\
\xi _{\mu ,0}^{(Mp)} &=&\xi _{e,0}^{(Mn)}\left( \frac{C_{\mu }m_{p}^{\ast }}{%
C_{e}m_{n}^{\ast }}\right) ^{2}\frac{(3k_{F,p}+k_{F,\mu}-k_{F,n})^{2}}{%
8k_{F,p}k_{F,e}}\Theta _{p\mu },
\end{eqnarray}
with $\Theta _{pl}=1$ for $k_{F,n}<(k_{F,l}+3k_{F,p})$ and $\Theta
_{pl}=0$ otherwise. We use $\alpha _{n}=1.76-0.63(1.68$
fm$^{-1}/k_{F,n})^{2},\beta _{n}=0.68$ from Page et al. (2004) here.
The tedious control functions to measure the superfluid effect can
be seen in Haensel et al. (2001).

%%%%%%%%%%%%%%%%%%%%%%%%%%%%%%%%%%%%%%%%%%%%%%%%%%%%%%%%%%%%%%%%
\begin{figure}[htbp]
\begin{center}
\includegraphics[width=0.6\textwidth]{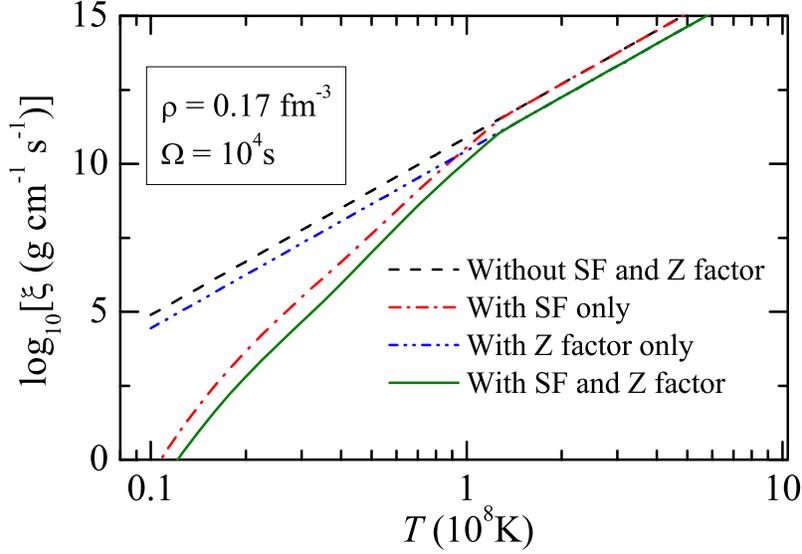}
\caption{(Color online) The logarithmic $\xi \text{ (g
cm}^{-1}\text{s}^{-1})$ under different temperature $T$ (in units of
$10^8$ K). The angular frequency is $\Omega=10^4$ s$^{-1}$. The
calculated results without the superfluidity (SF) and $Z$-factor,
with $Z$-factor only, with the $^3PF_2$ superfluidity only, and with
both the superfluidity and $Z$-factor, are shown for
comparison.}\label{fig:Xi}
\end{center}
\end{figure}
%%%%%%%%%%%%%%%%%%%%%%%%%%%%%%%%%%%%%%%%%%%%%%%%%%%%%%%%%%%%%%%%%%%%%%%

The calculated $\xi$ involving electron and muon branches as a
function of the temperature $T$, without and with the $Z$-factor and
its quenched superfluidity, are displayed in Fig.~\ref{fig:Xi},
taking the density $\rho=0.17$ fm$^{-3}$ and rotation angular
frequency $\Omega=10^4\text{ s}^{-1}$ as an example. The $Z$-factor
reduces the $\xi$ by about $50\%$ at $\rho=0.17$ fm$^{-3}$, and such
a $Z$-factor effect can be more substantial at high densities,
according to Eqs. (\ref{AA},\ref{MU1},\ref{MU2}). When the
temperature drops below the neutron $^3PF_2$ superfluid critical
temperature, the $\xi$ is significantly reduced. For instance, at
the temperature $T=10^7$ K, the $\xi$ is lowered by about six orders
of magnitude. The corresponding damping time scale $\tau_\xi$ is
thus enlarged especially at low temperatures. However, the bulk
viscosity dominates the damping mechanism just at high temperatures
($\gtrsim 10^{9}$K), at which the neutron $^3PF_2$ superfluidity
vanishes. It has been concluded that shear viscosity primarily stems
from electron scattering instead of neutron scattering (Shternin \&
Yakovlev 2008; Vidana 2012) at temperatures $T\gtrsim 10^{7}$ K.
Shang et al. found that the neutron shear viscosity $\eta _{n}$ is
enhanced by the $Z$-factor effect but is lowered much more strongly
by the onset of neutron triplet superfluidity. As a result, the
electron shear viscosity $\eta _{e}$ is generally larger than the
neutron one $\eta _{n}$ even at very low temperatures. So we still
employ the widely-used formulas for electron shear viscosity $\eta
_{e}$ (Shternin \& Yakovlev 2008; Alford et al. 2012) and
unimportant neutron shear viscosity $\eta _{n}$ (Cutler \& L.
Lindblom 1987) which are respectively given by
\begin{eqnarray}
\eta _{e} &=&4\times 10^{-26}\left( Y_{p}n_b \right)
^{14/9}T^{-5/3}\text{
\ g cm}^{-1}\text{s}^{-1}, \\
\eta _{n} &=&2\times 10^{18}\rho _{15}^{9/4}T_{9}^{-2}\text{ \ g cm}^{-1}%
\text{s}^{-1},
\end{eqnarray}
where $\rho _{15}$ and $T_9$ are the density and temperature in
units of $10^{15}$ g cm$^{-3}$ and $10^9$ K, respectively. $n_b$ is
baryon number density in units of cm$^{-3}$ here.

The time scales for bulk viscosity (Lindblom et al. 2002; Nayyar et
al. 2006), shear viscosity (Lindblom et al. 1998), and a $r$-mode
growth due to the gravitation radiation (Lindblom et al. 1998) are
respectively given by
\begin{eqnarray}
\frac{1}{\tau _{\xi }} &=&\frac{4\pi }{690}\left( \frac{\Omega ^{2}}{\pi G%
\overline{\rho }}\right) ^{2}R^{2l-2}\left( \int_{0}^{R}\rho
r^{2l+2}dr\right) ^{-1}\int_{0}^{R}\xi \left( \frac{r}{R}\right)
^{6}\left[
1+0.86\left( \frac{r}{R}\right) ^{2}\right] r^{2}dr, \\
\frac{1}{\tau _{\eta }} &=&(l-1)(2l+1)\left( \int_{0}^{R}\rho
r^{2l+2}dr\right) ^{-1}\int_{0}^{R}\eta r^{2l}dr, \\
\frac{1}{\tau _{\text{GW}}} &=&\frac{32\pi G\Omega ^{2l+2}}{c^{2l+3}}\frac{%
(l-1)^{2l}}{\left[ (2l+1)!!\right] ^{2}}\left(
\frac{l+2}{l+1}\right) ^{2l+2}\left( \int_{0}^{R}\rho
r^{2l+2}dr\right),
\end{eqnarray}
where $\overline{\rho }$ is the average density and $G$ is the
gravitational constant. The viscous dissipation at the viscous
boundary layer (VBL) of perfectly rigid crust and fluid core is
proposed as the primary damping mechanism in some literatures, and
its time scale is (Bildsten \& Ushomirsky 2000; Lindblom et
al. 2000)
\begin{equation}
\tau _{\text{VBL}}=\frac{1}{2\Omega }\frac{2^{l+3/2}(l+1)!}{l(2l+1)!!I_{l}}%
\sqrt{\frac{2\Omega R_{c}^{2}\rho _{c}}{\eta _{c}}}\int_{0}^{R_{c}}\frac{%
\rho }{\rho _{c}}\left( \frac{r}{R_{c}}\right)
^{2l+2}\frac{dr}{R_{c}},
\end{equation}
where $\rho _{c}$, $R_{c}$, and $\eta _{c}$ are the density, radius
and the shear viscosity of the NS matter at the outer edge
of the core (or equivalently inner edge of the crust). In the case
of $l=m=2$ mode, the $I_2$ is 0.80411 (Lindblom et al. 2000;
Rieutord 2001). Yet, a rigid crust is an ideal model. The relative
motion (slippage) between the crust and core reduces the damping by
as much as a factor of $f=10^2-10^3$ (Levin \& Ushomirsky 2001). In
the present work, we select $f=10^2$ as in the references of Ho et
al. (2011) with a slippage factor $\mathcal{S}=0.1$. Such a role is
questioned if the core-crust boundary is defined by a continuous
transition from non-uniform matter to uniform matter through
`nuclear pasta' phases with different shapes (Pethick \& Potekhin
1998), and consequently the VBL is smeared out (Gearheart et al.
2011).

An overall time scale of the $r$-mode, which includes the
exponential growth induced by the Chandrasekhar-Friedmann-Schutz
mechanism and the decay due to viscous damping, is given by $1/\tau
=-1/\tau _{\text{GW}}+1/\tau _{\xi }+1/\tau _{\eta }+1/f\tau
_{\text{VBL}}$. If $1/\tau>0$, the mode will exponentially grow,
while it will quickly decay if $1/\tau<0$. Therefore, a NS will be
stable against the $r$-mode instability if its angular velocity
$\Omega$ is smaller than a critical value $\Omega_c$. A star with
$\Omega > \Omega_c$ will lose its angular momentum through
gravitational radiation until the $\Omega$ falls below the $\Omega_c$.

\section{Effects of the $Z$-factor and the superfluidity on $r$-mode instability}\label{intro}\noindent

Usually the stellar surface temperature can be obtained from the
fitting of black-body spectra of LMXBs in quiescence, whereas the
core temperature cannot be obtained readily. Ho et al. (2012)
assumed that the stellar interior is isothermal and the heating is
balanced by neutrino cooling in steady state NSs of LMXBs, i.e.,
$L_{\text{heat}}=L_{\nu}$, to infer the stellar core temperature for
a NS with spin frequency $\nu_s=\Omega/2\pi$. The $L_{\text{heat}}$
is given by $L_{\text{heat}}=0.065(\nu _{s}/300$Hz$)L_{\text{acc}}$
(Brown \& Ushomirsky 2000), where $L_{\text{acc}}$ is the accretion
luminosity computed by using the observed flux and distance.
Usually, one assumes a canonical NS cools via the MUrca processes
when the core temperature $T \gtrsim 10^{8}$ K. Such a neutrino
emission process can be depressed by the neutron $^3PF_2$
superfluidity substantially when the core temperature is below the
critical temperature $T_c$. On the other hand, the superfluidity is
able to enhance the emission due to the Cooper pair breaking and
formation (PBF) process when the temperature is slightly below
$T_c$. This PBF process tends to be more effective than the MUrca
process, which could result in a rapid NS cooling (Page et al. 2011;
Shternin et al. 2011).

We calculate the neutrino emissivity $Q$ for $1.4M_{\odot}$
canonical NSs with the inclusion of the $Z$-factor and neutron
superfluid effects. The neutrino emissivity for DUrca, MUrca and PBF
processes are given as $Q^{(D)}=Z_{F,n}Z_{F,p}Q_{0}^{(D)}$,
$Q^{(Mn)}=Z_{F,n}^{3}Z_{F,p}Q_{0}^{(Mn)}$,
$Q^{(Mp)}=Z_{F,n}Z_{F,p}^{3}Q_{0}^{(Mp)}$ and $Q^{(PBF)}
=Z_{F,n}^{2}Q_{0}^{(PBF)}$ (Dong et al. 2016). Here $Q_{0}^{(D)}$,
$Q_{0}^{(Mn)}$, $Q_{0}^{(Mp)}$ and $Q_{0}^{(PBF)}$ are well-established emissivity that include the possible superfluid effect with
control functions (see Yakovlev et al. 1999, 2001 for details) but without the $Z$-factor
effect ($Z=1$). The
corresponding luminosity is calculated by volume integral $L=\int
QdV$. The NS interior is assumed to be composed of $npe\mu$ dense
matter without exotic degrees of freedom. The stellar structure for
a nonrotating NS is established by integrating the
Tolman-Oppenheimer-Volkov (TOV) equation with the EOS from the
user-friendly IMP1 Skyrme energy density functional (Dong \& Shang
2020). Such an EOS is close to the APR EOS because the APR EOS for
pure neutron matter has been served as a calibration for building up
the IMP1 Skyrme interaction (Dong \& Shang 2020), and provides good
description of neutron star properties. Within this EOS, the results
do not depend substantially on the assumed stellar mass.

%%%%%%%%%%%%%%%%%%%%%%%%%%%%%%%%%%%%%%%%%%%%%%%%%%%%%%%%%%%%%%%%
\begin{figure}[htbp]
\begin{center}
\includegraphics[width=0.55\textwidth]{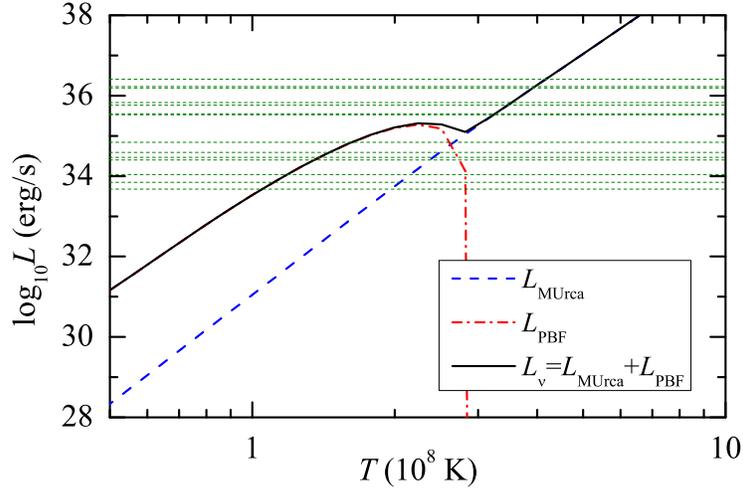}
\caption{(Color online) The calculated neutrino luminosity from the
MUrca, PBF processes for canonical NSs, where the $Z$-factor effect
and its quenched neutron $^3PF_2$ superfluidity are included. The
stellar structure is constructed based on the EOS from the IMP1
Skyrme interaction. The dashed horizontal lines denote the obtained
$L_{\text{heat}}$ for NSs in known LMXBs using the observed flux,
distance and spin frequency (Watts et al. 2008). The intersection of
the $L_{\text{heat}}=L_{\nu}$ yields the stellar core temperature
for each NS.}\label{fig:Lum}
\end{center}
\end{figure}
%%%%%%%%%%%%%%%%%%%%%%%%%%%%%%%%%%%%%%%%%%%%%%%%%%%%%%%%%%%%%%%%%%%%%%%

The calculated luminosity $L_{\text{MUrca}}$, $L_{\text{PBF}}$ and
the total neutrino luminosity
$L_{\nu}=L_{\text{MUrca}}+L_{\text{PBF}}$ are presented in
Fig.~\ref{fig:Lum}. At temperatures lower than the superfluid
critical temperature $T_c$, $L_{\text{PBF}}$ is about $2 \sim 3$
orders of magnitude larger than the $L_{\text{MUrca}}$, resulting in
a lower inferred core temperature $T$. The observed flux, distance
and spin frequency from Watts et al. (2008) are applied to achieve
the $L_{\text{heat}}$ for NSs in known LMXBs. The intersection of
the curve $L_{\nu}$ and $L_{\text{heat}}$ gives the stellar core
temperature for each NS. Unlike in Ho et al. (2011a, 2011b), each NS
we discussed here has a unique inferred core temperature, and it
lies in the range of $(1 \sim 3)\times 10^{8}$ K.

%%%%%%%%%%%%%%%%%%%%%%%%%%%%%%%%%%%%%%%%%%%%%%%%%%%%%%%%%%%%%%%%
\begin{figure}[htbp]
\begin{center}
\includegraphics[width=0.55\textwidth]{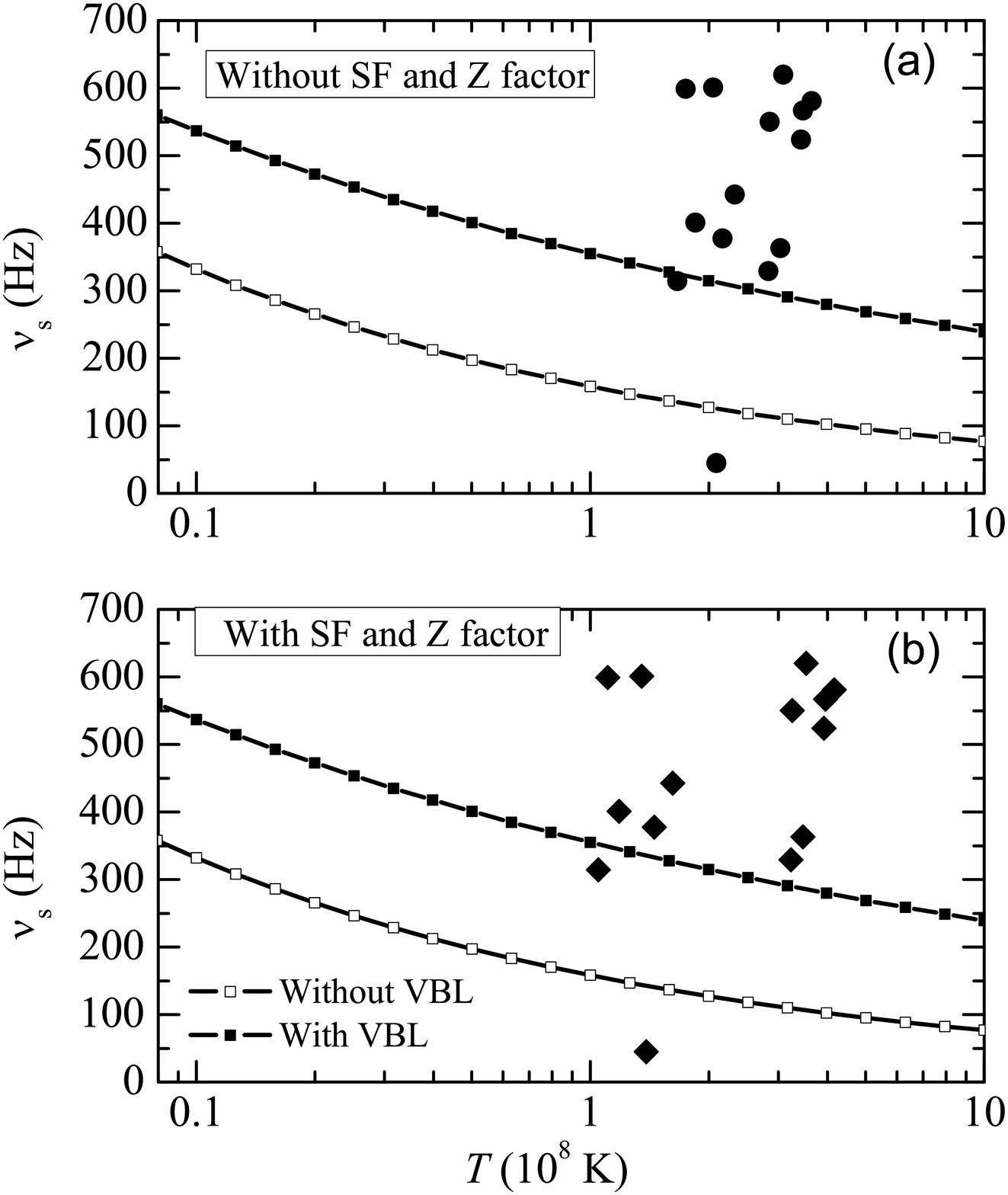}
\caption{The observed spin frequency $\nu_s$ and the inferred core
temperature $T$ for NSs in known LMXBs. The hallow squares do not
include the VBL damping, namely $1/\tau _{\text{GW}}=1/\tau _{\xi
}+1/\tau _{\eta }$. The black squares include the VBL damping,
namely $1/\tau _{\text{GW}}=1/\tau _{\xi }+1/\tau _{\eta }+1/f\tau
_{\text{VBL}}$, with $f=100$. Above the critical curves are the
$r$-mode instability region for $M = 1.4M_{\odot }$ canonical NSs
without (a) and with (b) the $Z$-factor and its quenched
superfluidity (SF). }\label{fig:r_mode}
\end{center}
\end{figure}
%%%%%%%%%%%%%%%%%%%%%%%%%%%%%%%%%%%%%%%%%%%%%%%%%%%%%%%%%%%%%%%%%%%%%%%

Figure~\ref{fig:r_mode} exhibits the $\nu_s-T$ plot where core
temperatures $T$ are inferred from $L_{\nu}=L_{\text{heat}}$ and
spin frequency $\nu_s$ have been observed for NSs in these LMXBs.
The error bar on the inferred $T$ is not included here because it
does not affect the following discussions. The $r$-mode instability
window is determined by the time scale balance $1/\tau
_{\text{GW}}=1/\tau _{\xi }+1/\tau _{\eta }+1/f\tau _{\text{VBL}}$.
This critical curve is almost not influenced by the $Z$-factor and
superfluid effects because the electron shear viscosity contributes
mainly to the viscous damping in this temperature range. When the
$Z$-factor and superfluidity are excluded, most of these millisecond
pulsars are located in the instability window, as shown in
Fig.~\ref{fig:r_mode}(a). When these two effects are included, the
referred core temperature of some rapidly rotating NSs are reduced
by a factor of around 2, as displayed in Fig.~\ref{fig:r_mode}(b),
being attributed to an enhanced neutrino emissivity from PBF process
(the $Z$-factor itself does not contribute distinctly).
Nevertheless, it is not evident to modify the conclusion that lots
of systems locate inside the unstable region, even if the damping
due to the VBL is included. So far, it is not fully understood what
causes LMXBs to undergo long versus short recurrence time bursts.
Intriguingly, as shown in Fig.~\ref{fig:r_mode}(b), these NSs are
clearly divided into two categories: high-temperature branch ($\sim
3\times10^8$ K) and low-temperature branch ($\sim10^8$ K). The short
LMXBs perhaps have high core temperature ($\sim3\times10^8$ K) than
long LMXBs ($\sim10^8$ K), as proposed by Ho et al. (2011). A higher
surface temperature is able to shorten the time intervals between
ignition of nuclear burning.

\begin{figure}[htbp]
\begin{center}
\includegraphics[width=0.55\textwidth]{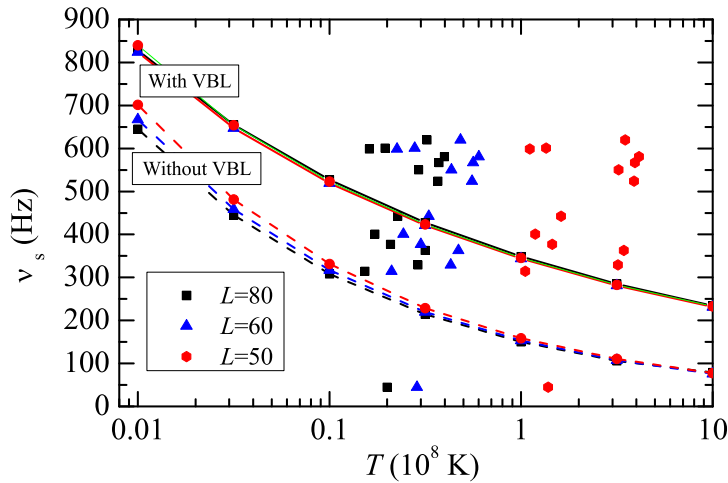}
\caption{(Color online) Similar to Fig. 4, but the NS masses are
assumed to be $M_{\text{TOV}} = 1.6M_{\odot }$. The EOS with various
density-dependent symmetry energy are employed, where the onset of
the DUrca process due to a stiff symmetry energy could lead to a
lower core temperature. The dashed (solid) curves are the critical
curves without (with) the VBL damping, and
$f=10^2$.}\label{fig:r_mode_L}
\end{center}
\end{figure}

The EOS used above is calculated by employing the IMP1 Skyrme force.
The symmetry energy are not stiff sufficiently to allow the onset of
the DUrca processes for canonical NSs and even large mass NSs. Yet,
recently Brown et al. (2018) showed that the NS in the transient
system MXB1659-29 has a core neutrino luminosity that is consistent
with the DUrca reaction occurring in a small fraction of the core,
substantially exceeds the MUrca processes, and they fitted the NS
mass $M\sim 1.6M_{\odot }$ in their model. We investigate the role
of the symmetry energy in the $\nu_s-T$ relation, and employ the
density-dependent symmetry energy written as
\begin{eqnarray}
S(\rho ) &=&13.0\left( \frac{\rho }{\rho _{0}}\right)
^{2/3}+C_{1}\left( \frac{\rho }{\rho _{0}}\right) +C_{2}\left(
\frac{\rho }{\rho _{0}}\right)
^{\gamma }, \\
C_{1} &=&19.4-\frac{18.3}{2.06-3\gamma \cdot 0.69^{\gamma }},  \notag \\
C_{2} &=&19.4-C_{1},  \notag
\end{eqnarray}
to supplement to the EOS of symmetric matter from IMP1 manually. It
can be considered as an extension of the DDM3Y-shape expression
(Mukhopadhyay et al. 2007; Dong et al. 2012; Dong et al. 2013; Fan
et al. 2014) and enables us to reproduce the symmetry energy
$S=32.4$ MeV at nuclear saturation density $\rho_0$ and the slope
parameter $L=42$ MeV at $\rho=0.11$ fm$^{-3}$ (Dong et al. 2018).
The only one free parameter $\gamma$ or equivalently the slope
parameter $L$ (distinguished from aforementioned luminosity $L$) at
$\rho_0$ controls its density-dependence (i.e., whether the symmetry
energy is stiff or soft). Vidana (2012) and Wen et al. (2012)
discussed the effects of symmetry energy on the $r$-mode
instability, which give opposite conclusions.
Figure~\ref{fig:r_mode_L} presents the calculated $r$-mode
instability critical curve and the location of each NS with
$M_{\text{TOV}}= 1.6 M_{\odot }$ in LMXBs in the $\nu_s-T$ plot,
taking $L=50, 60$ and 80 MeV as examples. If the DUrca process
opens, such as the $L=60$ MeV case, the inferred core $T$ is reduced
by about one order of magnitude because of the high efficiency of
the DUrca reaction. Although the neutrino emissivity of the DUrca
reaction can be about eight orders of magnitude larger than that of
the MUrca process, its influence on the inferred core temperature is
only about one order of magnitude because of the strong temperature
scaling in the $L_\nu-T$ relation. Many NSs are inside the stable
region, and others are also located in but closer to the stability
window. Therefore, the presence of the DUrca process could alleviate
the disagreement between the observed and the predicted results to a
large extent. Yet, if the VBL is smeared out because of the nuclear
pasta phases, the DUrca process still cannot modify the conclusion
that most NSs locate well inside the unstable region.

Additional damping mechanisms or physics perhaps is required to
reconcile the theory and observations. The mutual friction due to
vortices in a rotating superfluid (electrons scattered off of
magnetized vortices) is claimed to be unlikely to suppress the
$r$-mode instability in rapidly spinning NSs, but for a large `drag'
parameter $\mathcal{R}$ the mutual friction is sufficiently strong
to suppress this instability completely as soon as the core becomes
superfluid (Haskell et al. 2009). Anyway, this vortex-mediated
mutual friction damping is a rather complicated mechanism that is
difficult to calculate precisely, calling for further investigation.
In addition, the hyperonic bulk viscosity caused by nonleptonic weak
interactions is also found to stabilize the oscillation mode
effectively (Jha et al. 2010). Gusakov et al. (2014) argued that
finite temperature effects in the superfluid core leads to a
resonance coupling and enhanced damping of oscillation modes at
certain temperatures, and therefore the rapidly rotating NSs may
spend a long time at these resonance temperatures. Moveover, the
theoretical understanding can be consistent with observations if the
$r$-mode saturation amplitude is so small that the gravitational
wave torque cannot counteract the accretion torque although $r$-mode
heating is balanced by stellar cooling. As a result, the $r$-mode
instability has no impact on the spin or thermal evolution of NSs.
The recent investigations indeed present low saturation amplitudes
$\alpha_m$. For instance, Haskell \& Patruno (2017) constrained the
amplitude of an unstable $r$-mode from the spinning down of PSR
J1023+0038, and gave $\alpha_m \approx 5\times 10^{-8}$. Haskell et
al. (2012) concluded that for most known LMXBs in the unstable
region one has $\alpha_m=10^{-9} \sim 10^{-8}$. Using known X-ray
upper bounds on the temperatures and luminosities of several
non-accreting millisecond radio pulsars, Schwenzer et al. (2017)
derived the $r$-mode amplitude as low as $\alpha_m \lesssim
 10^{-8}$.

%%%%%%%%%%%%%%%%%%%%%%%%%%%%%%%%%%%%%%%%%%%%%%%%%%%%%%%%%%%%%%%%
%%%%%%%%%%%%%%%%%%%%%%%%%%%%%%%%%%%%%%%%%%%%%%%%%%%%%%%%%%%%%%%%%%%%%%%

\section{Summary}\label{intro}\noindent

We have investigated the bulk viscosity $\xi$ and $r$-mode
instability under the influence of both the $Z$-factor and its
quenched neutron superfluidity, where the recently calculated
superfluid gaps and $Z$-factors at Fermi surfaces within a
microscopic nuclear many-body approaches are employed. The
$Z$-factor effect reduces the bulk viscosity $\xi$ by several times at most,
while the neutron $^3PF_2$ superfluidity is able to reduce the $\xi$
by several orders of magnitude when the core temperature $T$ is
lower than the critical temperature. With the inclusion of
superfluidity, the PBF process opens when the stellar core
temperature is slightly lower than the critical temperature, leading
to an enhanced neutrino emission. As a result, the superfluidity
decreases the inferred core temperature for some relatively low
temperature NSs in LMXBs obviously. Interestingly, because of the
neutron $^3PF_2$ superfluidity, the core temperature of the NSs in
these discussed LMXBs are divided into two groups-high and low
temperatures. These NSs with long recurrence times for
nuclear-powered bursts are considered to have lower core temperature
($10^8$ K), while these with short recurrence times have high core
temperatures ($3\times10^8$ K). However, most NSs are still
predicted to be $r$-mode instable. In other words, the introducing
of the $Z$-factor and neutron triplet superfluidity cannot solve
this problem fundamentally. If the DUrca process occurs due to a
sufficiently stiff symmetry energy, the inferred core temperature is
reduced by about one order of magnitude, many NSs are located in the
$r$-mode stable window and others are closer to this region if the
VBL damping is taken into account. In other words, more NSs will be
inside the $r$-mode stability window for interactions which give
larger values of symmetry energy slope $L$. However, the existence
of most rapidly rotating NSs, such as the 716 Hz PSR J1748-2446 ad,
remains a puzzle. If the $r$-mode saturation amplitude is too small
to impact on the spin or thermal evolution of NSs, such as the
inferred $\alpha \approx 5\times 10^{-8}$ from the spinning down of
PSR J1023+0038 by Haskell \& Patruno (2017), the theoretical
understanding can be consistent with observations.

\section*{Acknowledgement}

\label{intro}\noindent J. M. Dong would like to thank L. J. Wang for
helpful suggestions. This work was supported by the National Natural
Science Foundation of China under Grant No. 11775276, by the
Strategic Priority Research Program of Chinese Academy of Sciences under
Grant No. XDB34000000, by the Youth Innovation Promotion Association
of Chinese Academy of Sciences under Grant No. Y201871.

\section*{Data Availability}
The data used to support the findings of this study are available
from the corresponding author upon request.

%\end{CJK*}

\end{document}